# Global Population Growth as Socio-Economic Soft Matter System Dynamics Evolution

Agata Angelika Rzoska[1] and Aleksandra Drozd-Rzoska[2]

[1]University of Economics , Dept. of Marketing, ul. 1 Maja 50, 40-257 Katowice, Poland

[2]Institute of High Pressure Physics Polish Academy of Sciences,

ul. Sokołowska 29/37, 01-142-Warsaw, Poland

[1]ORCID: 0000-0003-4121-31521 ;  e-mail: agatka.angelika@gmail.com

e-mail: agata.rzoska@edu.uekat.pl

[2]ORCID: 0000-0001-8510-2388;  e-mail: ola.drozdrzoska@gmail.com

 e-mail: arzoska@unipress.waw.pl

Abstract

The report considers the dynamics of the global population as the unique case of the Socio-Economic Soft Matter system. This category was introduced for complex systems dominated by mesoscale assemblies, emerging due to the inherent tendency for local self-organization. The hypothesis is validated by studying population growth evolution using universalistic scaling patterns developed in Soft Matter science. It is supported by the innovative derivative-based and distortions-sensitive analysis, showing the extended Malthus-type trend from 10 000 B till ca. the year 1200. Subsequently, the explicit evidence of the powered exponential population rise pattern is shown, with the unique crossover near 1970. Following this year, the population growth systematically slows down compared to earlier trends.

Population growth is confronted with global food demand evolution, which changes and also follows an exponential pattern. The rise of networking and innovations are indicated as the driving force leading to the crossover from the Malthus-type exponential behavior to the powered exponential one. It is supported by the analysis of the number of patents for innovations. The authors introduced the derivative-based and distortions-sensitive analysis for the optimal implementation of the powered exponential function for describing dynamic data.

# 1. Introduction

Soft Matter category introduced by Pierre G. de Gennes in his 1991 Nobel Prize lecture [1]. He was motivated by a set of universalistic features observed in so different physico-chemical systems as critical liquids, liquid crystals, polymers, micellar systems, and bio-systems – including living/active matter such as assemblies of bacteria or viruses, and food as the case of very complex Soft Matter [2-9]. Recently, the Topological [10] and Quantum Soft Matter [11] have been included.

This report indicates that the Global Population can be considered as the Socio-Economic Soft Matter parallel. It is shown that Global Population Growth (GPG) can be effectively described using scaling patterns developed for dynamics in Soft Matter Systems. It is supported by the innovative distortions-sensitive and derivative-based analysis, revealing GPG features hidden so far.

'Soft Matter' can be defined via a set of universalistic features shared amongst systems mentioned above, namely [1-11]:

**(i)** the dominance of mesoscale, macromolecular, or multimolecular (multi-particle) collective ' assemblies, often self-organized

**(ii)** the richness of phase transitions, i.e., qualitative changes of system symmetry-related features, at the local- and macro- levels when shifting temperature, pressure,… or during the temporal evolution

**(iii)** complex dynamics, which refers to time-dependent properties

**(iv)** unusual sensitivity to perturbations, both exo- and endogenic.

One can consider the following parallels for the hypothetical Socio-Economic Soft Matter parallel:

**ad. (i)** <u>the domination of structures in the mesoscale</u> can means the tendency to organize into tribes, cities, and states or organizations (political, trade-focused,…) networking different units.

**ad. (ii)** <u>richness of phase transitions</u>: revolutions or economic crises, with common 'pretransitional effects', leading to qualitative socio-economic changes

**ad. (iii)** <u>complex dynamics</u>: specific time evolutions of economic and social processes seem to be their intrinsic feature

**ad. (iv)** high <u>sensitivity to perturbations:</u> such as climate changes, energy and resources supply problems, or political problem: both external (exogenic) and internal- (endogenic) originated.

The focus on the global scale can facilitate the emergence of Socio-Economic Soft Matter. The action of agents that can disturb smaller units (tribes, cities, states,…) cease to be significant. They are averaged and finally appear as a kind of mean-field [12] acting on all elements of the system. It minimizes the number of relevant parameters and simplifies the model analysis,

The evolution of time-related (i.e., dynamic) properties in Soft Matter systems is most often described by the following scaling equations: 2, 5, 7, 12-16]:

$$P(t) = p_0 exp\left(\frac{\Delta t}{\tau}\right) \qquad (1)$$

$$P(t) = p_0 exp\left(\frac{\Delta t}{\tau}\right)^{\beta} \qquad (2)$$

$$P(t) = p_0 (\Delta t_C)^x \qquad (3)$$

where $P$ denotes the tested magnitude, $p_o$ is for the prefactor, $\tau_i$ is the processing time constant, $\beta$ is the power exponent; $\Delta t = t - t_{ref.}$, and $t_{ref.}$ is for the reference (onset) time; $\Delta t_C = t - t_C$ denotes the distance from some 'future critical singularity'.

Generally, in Eq. (1) the relaxation time denotes the time required to change the magnitude $P(t)$ by the factor $1/e \approx 0.3678. ...$, developed since the onset time. The convenient metric can be the value $\tau_{1/2} = \tau \times ln2$ describing the time required for 50% change.

The explanation of the powered exponential Eq. (2) is more complex. For $\beta > 1$ the ' compressing of relaxation processes occurs. It is hardly observable in physico-chemical systems case, except $\beta = 2$, which is equivalent to the normal (Gaussian) distribution [2], for a random variable governed processes. The stretched-exponential (SE) behavior related to (<1, associated with the distribution of relaxation times, is often observed in Soft Matter complex systems [7, 12-16]. Such behavior occurs in vitrifying or critical liquids, for instance [2-9]. So far, the analysis of dynamic 'experimental' data by SE Eq. (2) is carried out using the nonlinear fitting routines, which introduces notable uncertainty. For $\beta = 1$, Eq. (2) is reduced to the simple exponential Eq. (1). It is the case of a single relaxation time behavior in physical chemistry called the Debye process [2, 4, 12].

The power-type, non-exponential Eq. (3) is related to systems with an inherent critical singularity. This kind of deterministic attractor seems to be beyond the Socio-Economic Soft Matter parallel.

The report considers the dynamics of the global population as the unique case of a Socio-Economic Soft Matter system. The hypothesis is validated by testing the ability of the above scaling relation to describe the global population growth and the evolution of related properties. The analysis is supported by the innovative distortions sensitive analysis. It also solved the long-standing problem regarding the analytic implementation of SE Eq. (2) for describing 'experimental data' in complex systems.

## 2. The relation to classic population growth models

The onset of the modern population growth studies can be linked to Robert Malthus, who published '*An Essey on the Principle of Population*' in 1798. It can be encountered in the most influential publications in history, remaining a significant reference till nowadays. Malthus suggested the exponential growth of the human population (*P*) [17, 18]:

$$P(t) = p_0 exp(r \times t) \tag{4}$$

where the parameter $r = const$ is called the Malthusian parameter of population growth, or intrinsic rate of increase.

He confronted the exponential population rise with the assumed linear dependence of food resources growth and pointed to the inevitable crisis of a growing population. He proposes 'solutions' for impeding dilemma crisis, overtly non-ethical for nowadays standards. Trivial and politically convenient conclusions from the Malthusian model have been broadly used to justify war or colonial activities through plundering resource augmentation justified by the greater need to avoid crisis and enable development - at the expense of other "less important" nations. Unfortunately, the echoes of such 'imperial' behavior are also revealed today as the 'justifications' for the invasion of Ukraine. Pierre F. Velhulst [19, 20], considered development when the amount of available resources is limited, and there is no way to its supplementation. He showed the rapid Malthus-like population growth, followed by stagnation with the population remaining constant. Velhulst model is often recalled in microbiology for describing the evolution of bacterial colonies in a container with an initially determined amount of food. They develop following Velhulst model, but in the final stage, their number diminishes until their total death [21]. The question arises if such a model can be implemented for any human population. In the authors' opinion, it could occur in Rapa Nui (Easter) Island. After Polynesian sailors settled them, there was a sharp increase in the population and complete deforestation of a small and isolated island. The latter means well remote Pacific islands and the South America mainland. Deforestation was greatly exacerbated by the construction of about 1,000 enormous great Moai statues. The lack of wood meant there was no way to build a boat, so they were cut off from the abundant food resources in the ocean waters surrounding the island. This first caused the stabilization of the population and then its rapid decline until reaching the low level adjusted to remaining meager resources. In the decay phase, fraternal fights completely destroyed the grand culture of the Island. In March 1774, captain James Cook found only a small vegetating population [22]. From Malthus and Velhulst times, numerous population models have appeared. Generally they are arranged as Malthusian and non-Malthusian type. They take into account numerous socio-economic and natural factors, omitted in the Malthus and Velhulst models. [23-30] These models are recalled in numerous review reports and will not be discussed in this work.

Notwithstanding, the prediction of population growth remains a challenge. For instance, United Nations official prediction of the global population ranges from 8.5 billion to 12.5 billion in the year 2100. [27] The reliable prediction of the population growth is necessary for the strategic plans regarding infrastructure, resources, or the projection of possible solving socio-economic problems. In Modern times global scale forecasts have become crucial. In the 21st century, Earth has become a single organism, highly interactive at local and medium levels.

This report, as the first functional reference for the global population growth, considers the simple exponential Eq. (1), which from the Malthus Eq.(4) differs only by explicit taking into account a carefully selected reference onset time $t_{ref.}$. It was linked to the dawn of the Holocene epoch, nowadays also recalled as the Anthropocene.

## 2    Results and Discussion

10 000 BC, the Ice Age terminated, and the 'warm' Holocene epoch began. Between 16 000 BC and 10 000 BC, the global average temperature rose by 4.5 K. Before 10 000 BC, people lived in small and weakly interacting groups, which focused on survival in harsh environmental conditions [31]. In Holocene, humans could act beyond the daily struggle to survive. At 9,000 BC, the Neolithic time began [31, 32]: humans developed new generations of stone tools, more and more specialized. The first significant 'architectonic structures' appeared [31], the enormity of which leaves no doubt that they required the advanced cooperation of huge teams and interactive planning. The Stonehenge monument [32] can serve as the best example. It was possible due to groups, tribes, pre-villages, and pre-cities…spontaneously emerging, internally organized, and interacting.

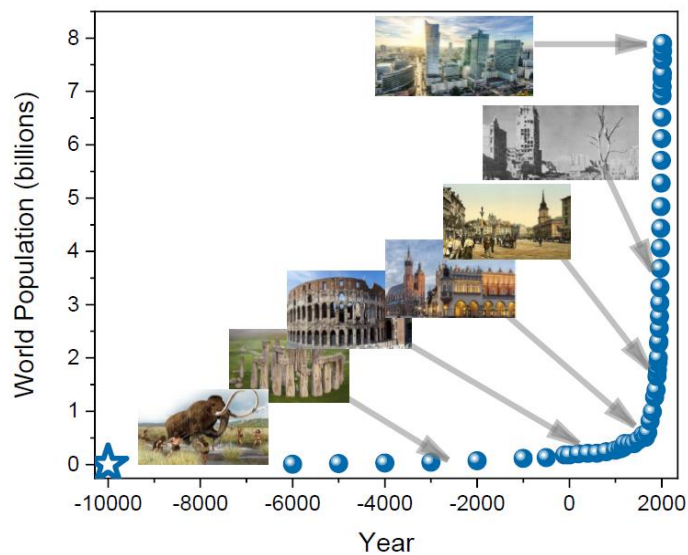

**Fig. 1** The 'classic' plot of the world population growth from ancient times until now (2021). The negative sign is related to BC (BCE) period. Pictures illustrating passing human epochs are included. They pertain to mammoth hunters, Stonehenge, the Colosseum in ancient Rome, medieval Cracow (Poland), Warsaw at the end of 19[th] Century (Poland), Warsaw in 1945 (the end of WW II), and Warsaw nowadays. Based on data from refs. [33-36].

All of these resemble the spontaneous emergence of collective mesoscale structures in Soft Matter systems when changing 'environmental' conditions (temperature, pressure, … for material systems). Figure 1 presents the classic chart used to illustrate the global population growth from the Holocene onset till nowadays, using population data from refs. [33-36].

The simplest way of parameterizing data, presented in Figure 1, offers the Malthus-type Eq. (1):

$$P(t) = p_0 exp\left(\frac{\Delta t}{\tau}\right) \quad \Rightarrow \quad lnP(t) = lnp_0 + \frac{\Delta t}{\tau} = lnp_0 + \lambda \Delta t \tag{5}$$

where: $\Delta t = t - t_{ref.}$.

As the reference $t_{ref.} = -12\,000\,BC$ was assumed. It was the dawn of the Holocene: the average global temperature rose by 2.5 K compared to the formal terminal of the Ice Age at 16 000 BC [31], and environmental restrictions for the human population qualitative weakened.

Eq. (1) linked to the simple differential equation:

$$\frac{dlnP(\Delta t)}{d\Delta t} = \frac{dlnP(\Delta t)}{dt} = \frac{1}{\tau} = \lambda = r = const \tag{6}$$

for $\tau = const$ in the given time period.

Parameters $\tau$ and $\lambda$ can be determined from the plot defined by Eq. 5, i.e., $lnP(t)$ or $log_{10}P(t)$ vs. $\Delta t$ or $t$, or numerically using Eq. (6). The analysis started from 5 000 BC, for which reliable enough population estimations are available. It revealed that the simple Malthus-type behavior (Eqs. (1 and 5) obey for millennia, from ca. the year 1 200 down to 10 000 BC. For the latter, the estimated population 1 750 000 inhabitants were obtained.

In Soft Matter science, the single exponential (Eqs. 1, 2) portrayal is the hallmark of weakly interacting species or their interplays reduce to the 'mean field interaction factor', approximately the same for all species [2, 12].

The year 1200 coincides with the border between the High- and Late-Middle Ages [37]. In the following times the population growth 'speeds up' and becomes 'nonlinear', for the semi-log scale used in Fig. 2. Near the year 1700, the slowly rising changes to a much 'faster' pattern. Analytically, such behavior is called the 'inflection point'. The year ~ 1700 can be associated with the beginning of the Enlightenment era in the sphere of ideas and the 1st Industrial Revolution [38]. A new socio-economic system emerged in Great Britain after the Cromwell revolution in the mid of 17th century. It allowed for the tremendous advancement of gifted people, necessary for developing the modern industry. Isaac Newton's ideas radiated and motivated. The emerging industry needed materials, preferably steel, and an energy source. Windmills, water mills, and mini-smelters based on mass forest exploration were used earlier. However, these 'technological solutions' were not satisfactory for growing needs at the onset of 18th century.

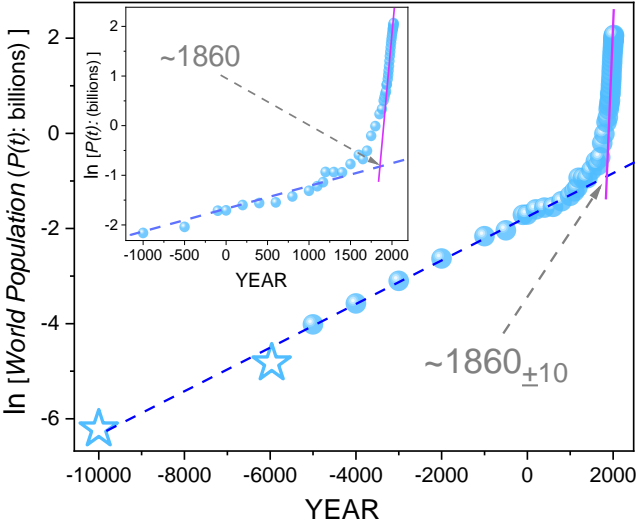

**Fig. 2** The Global human population growth from 10 000 BC till nowadays. The inset shows the focused behavior from the fall of the Roman Empire till nowadays. The crossover year where two dominant trends intersect is indicated. Lines are related to Eqs. (1) and (4). Parameters are given in Table I. Based on data from Fig. 1.

Suddenly, an innovative response appeared: coal as the energy source was discovered and, most importantly, broadly used in practice. But coal mines required machines for drainage, ventilation, and lifts, ... In 1710, Thomas Newcomen constructed the first steam engine, later qualitatively improved by James Watt , solving this fundamental problem. Thanks to coal, big amounts of 'innovative quality' steel could be produced. The production did not depend on the wood resources of surrounding forests.

The implementation of coal as the energy source created the grand transport problem of transporting massive & heavy goods (coal, iron ore, steel, and related products). The challenge was solved by the invention of the railroad and its 'immediate & broad' implementation and development (George Stephenson, 1814). It was possible due to the availability of large amounts of high-quality steel and coal. In the 19th century, the 1st Industrial Revolution

'covered' all of Europe and, subsequently, the World. It was the Steam Age [38, 39]. Finally, the apparent Malthus-type behavior (Eqs. 1 and 5) seems to emerge again when approaching Modern Times. The intersection between the 'millennial' and 'modern' Malthus-type trends occurs near the year 1860, as indicated in Fig. 2. It coincides the onset of Modern times, indicated for years 1830 – 1870, occurring due to the cumulation of technological and socio-economic changes [40]. The direct presentation of population growth data based on the semi-log scale (Eq. 5) can introduce averaging, hiding more subtle features. The distortions-sensitive and derivative-based can solve this functional problem: it offers an ultimate validation and can reveal hidden features. For the Malthus-type behavior the horizontal linear behavior should appear, as defined in Eq. (6). However, the plot the plot $dlnP(t)/dt$ vs. $t$ revealed disturbances from the Malthus-type behavior. One can take them into account by assuming temporal dependence of $\tau(t)$. $\lambda(t)$, what yields the extended Eq. (1):

$$P(t) = p_0 exp\left(\frac{\Delta t}{\tau(t)}\right) = p_0 exp(\Delta t \times \lambda(t)) \qquad (7)$$

Consequently, instead of simple Eq. (6), one obtains:

$$\frac{dlnP(t)}{dt} = \frac{d}{dt}exp\left(\frac{t-t_{ref.}}{\tau(T)}\right) = \frac{1}{\tau(t)} + \Delta t_{ref.}\frac{d}{dt}\left(\frac{1}{\tau(t)}\right) = \frac{1}{\tau(t)} - \frac{\Delta t}{(\tau(t))^2}\frac{d\tau(t)}{dt} = \lambda(t) \qquad (8)$$

The condition $dlnP(t)/dt = 0$ is related to an extremum of $P(t)$ or the inflection point, for which the slow-rise changes to the fast one. It is associated with the crossover: $\lambda(t) > 0 \longleftrightarrow \lambda(t) < 0$.

The preliminary analysis of $dlnP(t)/dt$ vs. $t$ confirmed the prevalence for the Malthus-type in the mentioned 11 000 years period, but with a few 'localized in time' disturbances. It means $\tau, \lambda = const$ for the general trend and local temporal distortion where $\tau, \lambda \to \tau(t), \lambda(t)$ correction should be taken into account. The most evident correlates with Roman Empires times. Starting from the late Medieval, the derivative-based plot reveals the almost permanent strong, strong and nonlinear rise of $dlnP(t)/dt$, which in practice excludes the basic Malthus-type evolution (Eqs. 1 and 4). As shown below, it can be functionalized using the powered exponential Eq. (2). However, its linearization faces an unconquerable problem so far, also in areas of physical chemistry and material engineering. Namely, the linearization of the powered exponential Eq. (2) has the form:

$$P(t) = p_0 exp\left(\frac{\Delta t}{\tau}\right)^\beta \Rightarrow lnP(t) = lnp_0 + \left(\frac{\Delta t}{\tau}\right)^\beta \qquad (9)$$

So far, the application of Eq. (9) required a nonlinear fitting routine and the knowledge of the domain in which it can be used, assumed apriori [2, 7, 12, 15]. The authors of this report solved this challenge: namely the derivative of Eq. (9) yields:

$$\frac{dlnP(t)}{dt} = \frac{\beta}{\tau}\left(\frac{\Delta t}{\tau}\right)^{\beta-1} \qquad (10)$$

Acting on both sides of the above relation by the logarithmic function, one obtains:

$$y(t) = log_{10}\left[\frac{dlnP(t)}{dt}\right] = log_{10}\left(\frac{\beta}{\tau}\right) + (\beta-1)log_{10}\frac{\Delta t}{\tau} =$$

$$= \left[log_{10}\left(\frac{\beta}{\tau}\right) - (\beta-1)log_{10}\tau\right] + (\beta-1)log_{10}\Delta t = A + B \times x \qquad (11)$$

The plot is based on transformed data as defined by Eq.(9), i.e., $y(t) = log_{10}[dlnP(t)/dt]$ vs. $x = log_{10}\Delta t$, should follow the linear dependence in the domain where the powered exponential description (Eqs. (2) and (9)) offers the optimal parameterization. The subsequent linear regression fit can yield values of relevant parameters.

Results of such analysis for the Global Population Growth are presented in Figures 3 and 4: first starting from the pre-antic times, and next since the year 1 800. Eq. (11) has been implemented using the logarithmic scale of 'y' and 'x' axis. It enables the discussion in the frame of Eq. (11), but also gives direct access to the simpler behavior defined by Eq. (7). The thick, dashed line in Fig. is for the dominant Malthus-type trend, extending from 10 000 BC. Notable are 'local, temporal' distortions in this domain from the mentioned dominant trend.

The most substantial disturbance is the period of the Roman Empire, which seems to lead to a specific population regression on a global scale.

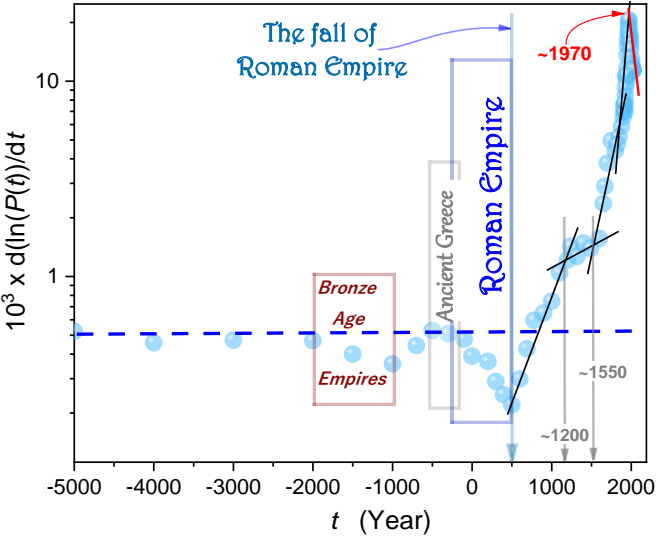

**Figure 3** The derivative-based and distortions sensitive analysis from ca. 5000 BC till nowadays. Significant historical periods and emerging dates are indicated. The dashed horizontal line is related to the basic Malthus-type behavior.

The influence of the Empire can be explained by the fact that even more than 1/5 of the Earth's inhabitants lived in it. Its creation motivated the emergence of highly organized structures in the neighborhood, such as the Parthian Empire. It was also the period of the final formation of the great Chinese Empire. However, the authors of the work were surprised by the detection of a growing population decline during the digestion of the Roman Empire, the achievements of which we are delighted to this day. Its achievements are considered a specific basis of modern civilization.[41] But great empires wage especially great and devastating wars. Relatively strong urbanization aided the catastrophic blows of disease plagues with the knowledge and possibilities of the ancient era. An inherent feature of all ancient empires, especially the Roman Empire, was the "exploitation" of enslaved people. The dignified life of every citizen of Rome was paid for by the death of numerous enslaved people: in mines, at great construction sites, thermal baths, or daily work in latifundia [41]. The end of the Roman Empire marked the beginning of non-Malthus-type population growth trends, which could mean better living conditions for the average person. Notable, that a hallmark of a similar decline from the dominated Malthus-type trend seems to appear between 2000 BC and 1000 BC, which correlates with times of great Bronze Age Imperia [41].

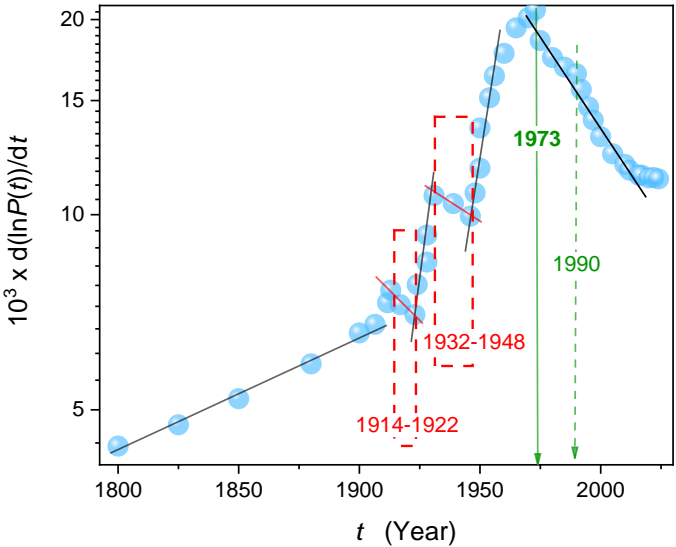

**Figure 4**  The derivative-based and distortions-sensitive analysis focused on the last two centuries, associated with Industrial Revolutions [38] defining Modern Times. Emerging characteristic dates are indicated.

After the fall of the Roman Empire the values $dlnP(t)/dt$ increases more than 40x, reaching the peak near the year 1970. As shown in Figs. 3 and 4 it is associated with a few linear sub-domains, i.e., the description by the powered exponential function (Eq. 1 and 9). Underlying historical events can influence them. They are briefly recalled in Figures. In 20$^{th}$ century the distortion associated with World War I extended from ca 1914 to 1922. Notable that World War II period only continued the population depression, which began near 1930 (Grand Crisis, Black Friday) and terminated in 1948. Worth emphasizing is the crossover occurring near the year 1970. Recalling that following Eq. (11) slopes of lines in Fig 3 and 4 determine the value of $\beta - 1$, one obtains the 'accelerated' to $\beta > 1$ powered exponential behavior before ~ 1970 and the stretched one related to $\beta < 1$ after 1970. It means that starting from the year 1970 the global population growth is weaker than one can expect from the earlier trend. The question arises of the reason for this unique change. Is it related to the end of the baby-boom after World War II ? May this is the consequence of the dramatic socio-cultural revolution in 1968? Concluding, the results emerging from the derivative-based and distortions-sensitive analysis, which results are shown in Figures 3 and 4, the following general function can be considered for portraying the global population growth:

$$P(t) = p_0 exp\left(\frac{\Delta t}{\tau(t)}\right)^{\beta} \qquad (12)$$

where $\tau(t)$ is the relaxation time, influenced by 'temporarily localized' even resulted from endogenic dynamics of population

There are two general domain described by Eq. (12)"

1. For a huge period reaching ~ 12,000 years, the dominant trend is description characterized by $\beta - 1 = 0$, i.e., the Malthus-type behavior coupled to $\beta = 1$. It can be temporarily distorted by endogenic, locally self-organized events, such as the appearance of the Roman Empires

2. From the late Middle Ages (> 1200) the explicit non-Malthus behavior associated with the exponent $\beta \neq 1$, emerges.

   The growth of the global population has to yield increasing worldwide demands for food. Figure 3 shows such a plot for the period in which a reliable, functional analysis is possible. The central part of the plot is for the 'classic, direct presentation'. The inset presents the same data analyzed via Eqs. (1) assuming the reference time $t_{ref.} = 1200$ year. Visible are two domains described by the mentioned relation: (1) for years between 1300 to 1700 and from 1930 till nowadays. The crossover between these trends can be estimated for the year ~1870. Characteristic is also the year 1700 when the inflection occurs. It can be associated with the beginning of the Enlightenment era in the sphere of ideas and the 1st Industrial Revolution [29]. A new socio-economic system emerged in Great Britain after the Cromwell revolution in the mid of 17th century. It allowed for the tremendous advancement of gifted people, necessary for developing the modern industry. Great ideas Isaac Newton's ideas radiated and motivated. The emerging industry needed materials, preferably steel, and an energy source.

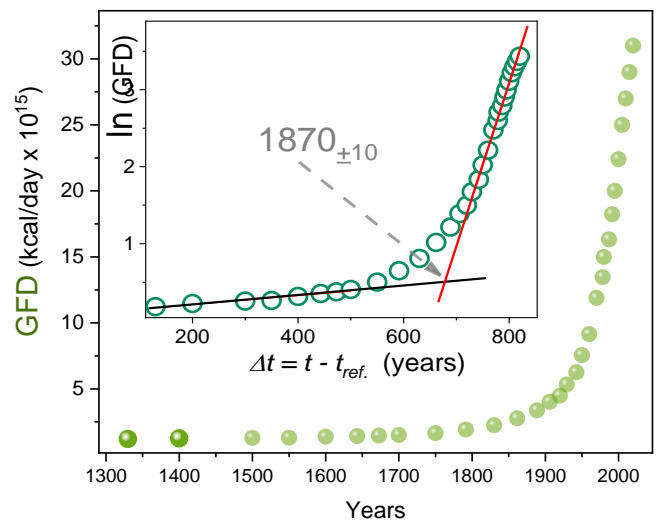

**Figure 5**     Global food demands & supply since Renaissance till nowadays. The inset shows the same data, taking the year 1200, as the reference. The central part of the plot is based on data from ref. [34].

Windmills, water mills, and mini-smelters based on mass forest exploration were used earlier. However, these 'technological solutions' were not satisfactory for growing needs at the onset of 18th century. Suddenly, an innovative response appeared: coal as the energy source was discovered and, most importantly, broadly used in practice. But coal mines required machines for drainage, ventilation, and lifts, ... In 1710, Thomas Newcomen constructed the first steam engine, later qualitatively improved by James Watt [39], solving this fundamental problem. Due to coal use, ample amounts of 'innovative quality' steel could be produced. It ceased to be limited by the wood resources available or not in surrounding forests.

Notwithstanding, the implementation of coal as the energy source created the grand transport problem of transporting massive & heavy goods (coal, iron ore, steel, and related products). The challenge was solved by the invention of the railroad and its 'immediate & broad' implementation and development (George Stephenson, 1814). It was possible due to the availability of large amounts of high-quality steel and coal. In the 19$^{th}$ century, the 1$^{st}$ Industrial Revolution 'covered' all of Europe and, subsequently, the World. It was the Steam Age [39].

In the authors' opinion, the driving force role for global development plays breakthrough innovation, scientific and technological, as well as in the sphere of ideas. Their feedback interactions with economic and social breakthroughs constitute another significant development factor. Figure 4 shows the global evolution of the number of patents related to innovation. +It is related to the period after 1890, which is associated with the 'firm establishing' the Patent Offices system on a global scale [42]. The central part of the figure shows the experimental data in the classic bar presentation, from which only a general conclusion can be drawn about the systematic and robust increase in the number of patents since ~ 1960. However, a more interesting picture emerges when analyzing these experimental data via Eqs. (1) and (4), with 1868 as the reference. It is shown in the inset in Fig. 4. Three linear domains validate the description via Eqs. (1) and (4). The first one extends from ~1890 to ~1930, which can be related to the *Grand Crisis* ('Black Friday'). Since then, there has been a continuous decrease in the number of patents, terminating in 1945. The immediate and robust rise occurred already in 1946. In the authors' opinion, this can be associated with the 'freezing' of inventors in the armies at war. From 1950 to the present day, a permanent increase in the number of innovations begins at a much higher level than in the pre-war period. The interesting anomaly appeared from the period 1982 - 1989, when the number of patents for innovations remained constant. Such an unusual behavior can be linked to the consequences of the 2$^{nd}$ oil crisis, associated with political changes in Iran. On the other hand, there is (almost) no hallmark of the 1$^{st}$ oil crisis (1973) impact. The mentioned period also correlates with the presidency of Ronald Reagan in the USA and Margaret Thatcher in the UK and Cold War intensification. In 1989, this peculiar innovation crisis terminated, which clearly correlates with the fall of communism in Eastern Europe.

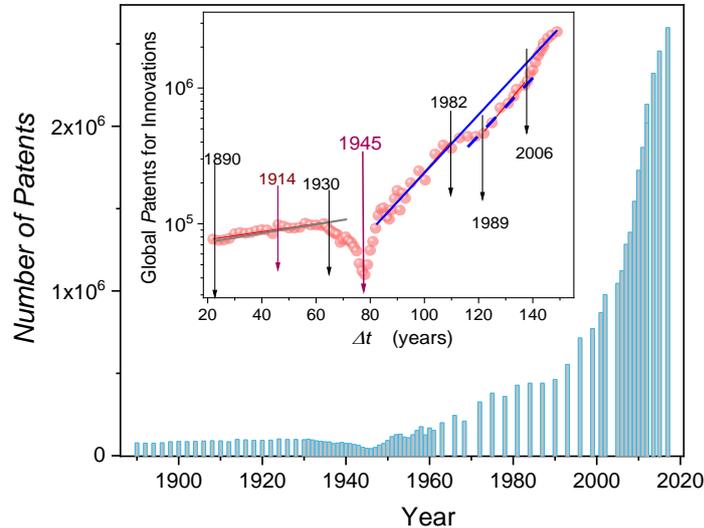

**Figure 6** Changes of the global number for innovation since the end of the 19$^{th}$ Century, till nowadays: 'classic' presentation. The inset shows the same data analyzed via Eqs. (1) and (4). with values of parameters given in Table I. Data for the central part of the plot from ref. [42].

The increase described by Eqs. (1) and (4) appears again. It is described with the same rate coefficient' ($r, \lambda$) value, but for the lower, 'shifted', the reference base (prefactor $p_0$ ). Notable that between the years 2006-2010 a substantial increase in the number of innovations occurred. The return to the 'higher' reference base occurred before the year 1982. Does it mean that the grand 'bank crisis', which happened at that times reset socio-economic surroundings to a more optimal pattern.

The author would like to stress that no massive famines occurred during the discussed 'innovation-driven' time, despite World War II. The Industrial Revolution [38] led to a particular increase in the scale of steel production, which inspired further innovations in its production and enabled the development and implementation of innovative solutions from the machine industry to construction and agriculture. Innovative types of equipment supporting the cultivation of plants appeared. Fertilizers and other innovations supported an extraordinary increase in plant yields, first thanks to guano imported from South America - which was possible thanks to coal-fed steel ships (steamers).

Later, it was produced industrially due to the extraordinary innovation of using nitrogen in the air [38]. Food storage and distribution are as necessary as food production, where losses could exceed 50%. This challenge solved innovations associated with developing chemical-preservation additives to food, enabling long-term storage without the risk of microbiological dangers and contaminations [33]. It was also the process of thermal pasteurization and sterilization. Innovations leading to the general use of refrigerators and food freezing. This boost of preservation-related innovations led to abundant and varied shelves with food in hypermarkets available to practically everyone. However, chemical preservatives also turned out to be the source of many disease-related problems. Thermal pasteurization radically reduces the bioactivity and amount of vitamines…. But again, an innovative answer to these problems has emerged in the last decades. The key example is the high-pressure preservation (HPP) for food, retaining the qualities of a fresh product for up to six months: without any chemical preservatives, including salt or sugar, or heat action [43-46].

## 3 Conclusions

This report shows that some large-scale, global, socio-economic issues can be discussed within general frames developed for the 'Soft Matter' science, characterized by the emergence of a common scaling pattern despite qualitative 'microscopic' (local) differences. Meaningful is only the spontaneous ability to appear and develop self-organized mesoscale structures. This path is tested for global population growth, revealing the existence of two domains, which can be portrayed by the powered exponential function, governed by the exponent $\beta$, with time-dependent relaxtion time or equivalently rate coefficient. The latter shows local, temporal, endogenic disturbance within the global complex population system. The first domain is for $\beta = 1$, i.e., which is equivalent to the extended Malthus- type behavior. It covers enormous period ~ 11 000 years and terminates near the year 1200. Later, the second domain governed by $\beta \neq 1$ emerges. Recalling the basics of Soft Matter science, such values can indicate the increasing complexity and networking within the systems. Near 1970 the crossover from the region governed by $\beta > 1$ to the region described by $\beta < 1$ occurs. It indicates that the population growth gradually slows down compared to the trend before 1970. Recalling Soft Matter science reference, it may indicate the stretched exponential behavior, associated with the broad distribution of relaxation times, and describe the broadening variety of processes. Finally, breakthrough innovations are indicated as the development driving force, with the qualitative renewal of crucial resources.

The authors would like to stress the innovative derivative-based and distortions-sensitive analysis for the optimal description of dynamic data by the powered exponential function.

It should be emphasized that the necessary breakthrough scientific and technological innovations may appear and affect development only if the appropriate economic and social innovations are also developed and implemented. Its task is the creation of a suitable socio-economic environment. The optimal development model is driven by the spontaneous feedback synergy for science-technological and socio-economic innovations, yielding a sustainable system.

**Competing interest statement:**

There are no competing interests for the authors

**Contribution:**

The authors declare equal contributions regarding all aspects of the paper.

**References**

1. P.G. de Gennes, J. Badoz, *Fragile Objects: Soft Matter, Hard Science, and the Thrill of Discovery*, Springer Verlag, New York, 1996
2. S.J. Rzoska, V. Mazur, A. Drozd-Rzoska (eds.), *Metastable Systems under Pressure,* Springer Verlag, Berlin, 2010

3. S.J. Rzoska, A. Drozd-Rzoska, P.K. Mukherjee, D.O. Lopez, J.C. Martinez-Garcia, *Distortion-sensitive insight into the pretransitional behavior of 4-n-octyloxy-4'-cyanobiphenyl (8OCB)*, J. Phys.: Condens. Matter **25**, 245105, 2013

4. S.J. Rzoska, A. Drozd-Rzoska, *Dual field nonlinear dielectric spectroscopy in a glass forming EPON 828 epoxy resin*, J. Phys.: Condens. Matter **24**, 035101, 2011

5. A. Drozd-Rzoska, S.J. Rzoska, J. Zioło, *Anomalous temperature behavior of nonlinear dielectric effect in supercooled nitrobenzene*, Phys. Rev. E **77**, 041501 (2008)

6. A. Drozd-Rzoska, S.J. Rzoska, J. Zioło, *Mean-field behaviour of the low frequency non-linear dielectric effect in the isotropic phase of nematic and smectic n-alkylcyanobiphenyls*, Liquid Crystals **21**, 273-277 (1996)

7. S.J. Rzoska, J. Zioło, A. Drozd-Rzoska, Stretched relaxation after switching off the strong electric field in a near-critical solution under high pressure, Phys. Rev. E 56, 2578 (1997)

8. A. Drozd-Rzoska, S.J. Rzoska, S. Pawlus, J. Zioło*, Complex dynamics of supercooling n-butylcyanobiphenyl (4CB)*, Phys. Rev. E **72**, 031501 (2005)

9. S.J. Rzoska, S. Starzonek, A. Drozd-Rzoska, K. Czupryński, K. Chmiel, .B. Szczypek, W. Walas, *Impact of nanoparticles on pretransitional effects in liquid crystalline dodecylcyanobiphenyl*, Phys. Rev. E **93**, 020701 (2016)

10. F. Serra, U. Tkalec, Lopez-Leon (eds), *Topological Soft Matter*, Front. in Physics **11**, 1-108 (2020).

11. R.P. Thedford, F. Yu,William R.T. Tait, K. Shastri, F. Monticone, U. Wiesner, *The Promise of Soft Matter Enabled Quantum Material*s, Adv.Mat. in press. https://doi.org/10.1002/adma.202203908 (2022).

12. S.J. Rzoska and V. Mazur (eds.), *Soft Matter under Exogenic Impacts*, NATO Sci. Series II, vol. 242, Springer, Berlin, 2007

13. A Drozd-Rzoska, S.J. Rzoska, S. Pawlus, J.C. Martinez-Garcia, J.L. Tamarit, *Evidence for critical-like behavior in ultraslowing glass-forming systems*, Phys. Rev. E **82**, 031501 (2010)

14. S.J. Rzoska, M. Górny, and J. Zioło, *Stretched-exponential relaxation of the nonlinear dielectric effect in a critical solution*, Phys. Rev. A **43** 1100 – 1102 (1991)

15. S.J. Rzoska, V. Degiorgio, T. Bellini and R. Piazza, *Relaxation of the electric birefringence near a critical consolute point*, Phys. Rev. E **49**, 3093-3096 (1994)

16. S.J. Rzoska, E. Rosiak, M. Rutkowska, A. Drozd-Rzoska, A. Wesołowska, M.K. Borszewska-Kornacka, *Comments on the high pressure preservation of human milk*, Food Sci. and Nutr. Studies **1**, 17 (2017)

17. T. Malthus, *An Essay on the Principle of Population,* Createspace Independent Publ. Platf., 1798 & 2013

18. D.N. Weil, J. Wilde, *How Relevant Is Malthus for Economic Development Today?*, Am. Econ. Rev. **100**, 378–382 (2010)

19. P.-F. Velhulst, *Recherches mathématiques sur la loi d'accroissement de la population*, Nouveaux Mémoires de l'Académie Royale des Sciences et Belles-Lettres de Bruxelles **18**, 1–42, (1845)

20. F.L. Ribeiro, *An attempt to unify some population growth models from first principles,* Rev. Bras. Ensino Fís. **39**, 1 (2017)

21. M. Peleg, M.G Corradini, M.D Normand, *The logistic (Verhulst) model for sigmoid microbial growth curve revisited*, Food. Res. Int. **40**, 808-818 (2007)

22. D.O'Leary, *Rapa Nui, Easter Island*; Floricanto Press, Puerto Rico, 2021

23. F. Cecconi ,M. Cencini M. Falcionia A. Vulpiani, *Predicting the future from the past: An old problem from a modern perspective*, Am. J. Phys. **80** , 1001-1008 (2012)

24. C. Bystroff, *Footprints to singularity: A global population model explains late 20th century slow-down, and predicts peak within ten years,* PLoS ONE **16**, e0247214 (2021)

25. A. van Witteloostuijn, J. Vanderstraeten, H. Slabbinck, M. Dejardin, J. Hermans, W. Coreynen, *From explanation of the past to prediction of the future: A comparative and predictive research design in the Social Sciences*, Soc. Sci. Hum. Open **6**, 100269 (2022)

26. E. Wesley F. Peterson, The Role of Population in Economic Growth

27. UN Population Division, Global Population Growth and Sustainable Development. Report (UN, NY, 2021)

28. E. Wesley F. Peterson, *The Role of Population in Economic Growth*, Sage Open **7**, 1-15 (2021)

29. A. Dias, M. D'Hombres, B. Ghisetti, C. Pontarollo, N. Dijkstra, *The Determinants of Population Growth: Literature review and empirical analysis*; Working Papers 2018-10, Joint Research Centre, European Commission, 2018.

30. D. Adam, *How far will global population rise? Researchers can't agree*, Nature **597**, 462-465 (2021)

31. N. Roberts, *The Holocene*, J. Wiley & Sons, New York, 2014

32. B. Cornwell, *Stonehenge*, HarperCollins Publ., New York, 2004

33. M. Roser, H. Ritchie, E. Ortiz-Espina, *World Population Growth,* https://ourworldindata.org/world-population-growth

34. https://ourworldindata.org/grapher/population

35. J. C. Caldwell and T. Schindlmayr, *Population and Development Review Historical Population Estimates: Unraveling the Consensus*, Popul. Devlop. Rev. **28**, 183-204 (2002)

36. . https://www.census.gov/data/tables/time-series/demo/international-programs/historical-est-worldpop.html

37. W. Blockmans, P. Hoppenbrouwers, *Introduction to Medieval Europe 300 -1500*, Routledge, New York, 2018

38. P.N. Stearns, *The Industrial Revolution in World History*, Routledge, New York, 2020

39. T. Crump, *A Brief History of the Age of Steam. The Power That Drove Industrial Revolution*, Robinson, London, 2007

40. K. Pobłocki, Dwie podróże po kapitalizmie, czyli odpowiedź Janowi Sowie (Two journeys around capitalism, or the response to Jan Sowa). *Praktyka Teoretyczna*, **26**, 387-410 (2018)

41. L. de Blois, R.J. van der Spek, *An Introduction to the Ancient World*, Routledge, New York, 2019

42. https://www.kilburnstrode.com/knowledge/european-ip/innovation-and-crisis

43. S. Thorne, *The History of Food Preservation*, Rowman&Littlefield Publ., London, 1986

44. T. Koutschma, *Adapting High Hydrostatic Pressure (HPP) for Food Processing Operations*, Elsevier, Amsterdam, 2014

45. B. Sokołowska, S. Skąpska, M. Fonberg-Broczek, J. Niezgoda, M. Chotkiewicz, A. Dekowska, S.J Rzoska, *Factors influencing the inactivation of Alicyclobacillus acidoterrestris spores exposed to high hydrostatic pressure in apple juice*, High Press. Res. **33**, 73-82 (2013)

46. A. Wesolowska, E. Sinkiewicz-Darol, O.Barbarska, K. Strom, M. Rutkowska, K. Karzel, Elzbieta Rosiak, G. Oledzka, M. Orczyk-Pawiłowicz, S.J. Rzoska, M.K. Borszewska-Kornacka, *New achievements in high-pressure processing to preserve human milk bioactivity*, Frontiers in Pediatrics **6**, 323 (2018)